\documentclass[preprint2]{aastex631}

\usepackage{newtxtext,newtxmath}
\usepackage{gensymb}

\submitjournal{ApJ}

\graphicspath{{./}{figures/}}
\begin{document}

\title[MHD nucleosynthesis]{Nucleosynthesis in the Innermost Ejecta of Magnetorotational Supernova Explosions in 3-dimensions}

\correspondingauthor{Shuai Zha}
\email{zhashuai@ynao.ac.cn}

\author[0000-0001-6773-7830]{Shuai Zha}
\affiliation{Yunnan Observatories, Chinese Academy of Sciences (CAS), Kunming 650216, China}
\affiliation{Key Laboratory for the Structure and Evolution of Celestial Objects, CAS, Kunming 650216, China}
\affiliation{International Centre of Supernovae, Yunnan Key Laboratory, Kunming 650216, China }

\author[0000-0002-4470-1277]{Bernhard M\"uller}
\affiliation{School of Physics and Astronomy, Monash University, Clayton, VIC 3800, Australia}

\author[0000-0002-1357-4164]{Jade Powell}
\affiliation{Centre for Astrophysics and Supercomputing, Swinburne University of Technology, Hawthorn, VIC 3122, Australia}
\affiliation{OzGrav: The ARC Centre of Excellence for Gravitational-Wave Discovery, Hawthorn, VIC 3122, Australia}

\begin{abstract}
Core-collapse supernova (CCSN) explosions powered by rotation and magnetic fields present an interesting astrophysical site for nucleosynthesis that potentially contributes to the production of $r$-process elements. Here we present yields of the innermost ejecta in 3D magnetorotational CCSN models simulated using the \textsc{CoCoNuT-FMT} code. Strong magnetic fields tap the rotational energy of the proto-neutron star and lead to earlier and more energetic ($\sim3\times10^{51}$\,erg) explosions than typical neutrino-driven CCSNe. Compared to a reference non-magnetic model, the ejecta in the magnetorotational models have much more neutron-rich components with $Y_\mathrm{e}$ down to $\sim0.25$. Our post-processing calculations with the reaction network \texttt{SkyNet} show significant production of weak $r$-process elements up to mass number $\sim$130. We find negligible differences in the synthesis of heavy elements between two magnetorotational models with different initial field strength of $10^{10}$ and $10^{12}$\,G, in accord with their similar explosion dynamics. The magnetorotational models produce about $\sim0.19$ and $0.14\,M_{\sun}$ of radioactive $^{56}$Ni, on the low end of inferred hypernova nickel masses. The yields are publicly available at Zenodo: doi:\href{https://doi.org/10.5281/zenodo.10578981}{10.5281/zenodo.10578981} for comparison with stellar abundance patterns, inclusion in modelling galactic chemical evolution, and comparison with other yield calculations. Our results add to the yet restricted corpus of nucleosynthesis yields from
3D magnetorotational supernova simulations and will help quantify yield uncertainties.
%and can push forward the understanding of its role in GCE. 
\end{abstract}


\section{Introduction} \label{sec:intro}

It has long been suggested that rotation and magnetic fields play an important role in some core-collapse supernova (CCSN) explosions of massive stars heavier than $\mathord{\sim}8\mathord{-}10\,M_{\sun}$ \citep{leblanc70,meier76}. In the modern view of magnetorotational (MR) CCSNe, strong magnetic fields amplified through magnetorotational instabilities (MRI) or some other form of dynamo action tap the rotational energy of the newly formed proto-neutron star (PNS) to launch bipolar jets and probably power a hyperenergetic explosion \citep{wheeler02,akiyama03,burrows07,dessart07,masada15,rembiasz16,kuroda20,martin20,martin21,bugli21,powell23,matsumoto23,shibagaki23}. The MR mechanism provides a plausible scenario for \emph{hypernovae} with an explosion energy of $\sim10^{52}\,$erg, about ten times that of normal CCSNe, which are often associated with long gamma-ray bursts (long GRBs, \citealt{galama98,bloom99,hjorth03}). Despite impressive progress in modelling MR explosions, the physical mechanism behind hypernovae and long GRBs is still not definitely settled and no single model can as yet explain the phenomenology of these extreme transients completely.
One lingering issue is, for example, that in several sophisticated 3D magnetohydrodynamic (MHD) simulations, the bipolar jets became less collimated or even destroyed at later times due to various non-axisymmetric instabilities (e.g., \citealt{mosta14,kuroda20,powell23}, but see \citealt{martin21} for stable jets). One complication is that the relativistic jets seen in long GRBs and the non-relativistic outflows that carry the bulk of the explosion energy are likely very distinct phenomena \citep{burrows07}. A unified understanding of the non-relativistic precursor to the GRB and the GRB itself will require models that bridge different phases and regimes in hypernova explosions to consistently account for multi-messenger observations of the observed events. Because of their yet enigmatic nature and the extreme energies involved, magnetorotationally-powered hypernovae are a major target for multi-messenger astronomy.
In particular, nearby energetic supernovae are important targets for current and future ground-based gravitational-wave observatories due to their apparent asphericity and unique emission features \citep{powell23,shibagaki23}, which might help constrain the explosion by an unobscured look into the supernova core.

With regard to multi-messenger astronomy,
another intriguing aspect of MR CCSNe is that they are a potential site for rapid-neutron-capture ($r$-process) nucleosynthesis \citep{nishimura15,nishimura17,mosta18,halevi18,winteler12}. MR explosions can be a more viable scenario than neutron-star mergers for explaining the \emph{early} enrichment of $r$-process elements in some extremely metal-poor stars (see \citealt{thielemann17} for an overview). \citet{yong21} reported an outstanding example, SMSS J200322.54-114203.3, with very low overall metallicity and significant $r$-process enrichment that can possibly be explained by pollution with the ejecta from a MR CCSN with a 25\,$M_{\sun}$ progenitor. The potential for $r$-process nucleosynthesis in MR CCSNe was already recognised before the advent of detailed simulations with magnetic fields and neutrino transport and has long been investigated based on yields of MR CCSNe by simulations with parameterized neutron-rich ejecta in the prompt-magnetic jet \citep{cameron03,nishimura15,grimmett21}. \citet{winteler12} and \citet{mosta18} took
significant leaps towards understanding neutron-rich nucleosynthesis in MR explosions using 3D MHD simulations with a neutrino leakage scheme to determine nucleosynthetic conditions.  The latest first-principle 3D MHD simulations together with spectral neutrino transport \citep{martin21,powell23,kuroda20} have made it possible to investigate whether MR CCSNe can provide the necessary conditions for the robust $r$-process production with much less uncertainty from parameterizations used in the past. Nucleosynthesis calculations based on the 3D neutrino-MHD model simulated with the \textsc{Aenus-Alcar} code \citep{martin21, reichert23} suggest $r$-process yields of MR CCSNe up to the 2nd peak, i.e., to mass number $A\sim130$. However, the results of \citet{reichert23,reichert24} also indicate non-trivial
model dependency of the yields.
More nucleosynthesis studies using different simulation codes, grid resolutions, and initial model setups are urgently needed to assess the robustness of nucleosynthesis yields from MR CCSNe and quantify uncertainties in yield predictions.

Sources of uncertainty in models of MR CCSNe are still manifold. One of the main uncertainties concerns the poorly understood evolution of massive stars with rapid rotation and magnetic fields. Currently, only 1D progenitor models are available with `shellular' rotation and prescriptions for rotation-induced mixing and magnetic fields \citep{spruit02,heger05,mesa13}. \citet{takahashi21} have made improvements in the modelling of the interplay between the magnetic fields and rotation by suggesting what is essentially a sophisticated physically-motivated turbulence model including rotation and magnetic fields, but their algorithm has not yet been used to evolve massive stars to collapse to provide the progenitor models required for CCSN simulation. Current stellar evolutionary simulations thus cannot confidently predict accurate magnetic field strength and multi-dimensional configuration at the onset of core collapse. This is relevant for MR CCSN nucleosynthesis studies, as \citet{reichert24} have noticed that the initial magnetic field configuration is an important factor determining the nucleosynthetic conditions of the inner ejecta in MR explosions. Note that recently 3D simulations of short phases of the pre-collapse progenitor evolution with magnetic fields have been performed for the non-rotating case \citep{varma21} and for a rapidly rotating case \citep{varma23}. These studies, however, only simulated a portion of the star (some active burning shells) and especially the rotating models did not cover sufficiently long secular timescales to provide consistent pre-collapse structures for CCSN simulations.
As another source of uncertainty, the post-collapse amplification of the magnetic fields may require very high numerical resolution to capture instabilities on small scales (e.g., 100\,m in \citealt{mosta15}), especially if the initial fields are weak. Such high resolution is not achievable in global CCSNe simulations with detailed neutrino transport. The common approach in MR-CCSNe modelling is therefore to impose relatively strong initial fields in the pre-collapse core so that the PNS can achieve magnetar-strength magnetic fields directly as a result of flux compression \citep{mosta14,martin21,powell23}. Interestingly though, the latest generation of 3D MHD supernova simulations is showing MR explosions even for more moderate initial field strengths of $\mathord{\sim}10^{10}\,\mathrm{G}$ that appear plausible in the light of recent 3D progenitor simulations.

Uncertainties also arise because of physical approximation and implementational differences in the numerical methods used in MR CCSN simulations. Neutrino interactions are crucial for determining the nucleosynthetic conditions, i.e. the neutron to proton ratio of the ejecta. Many simulations in the last decade \citep{mosta14,mosta18} employed a neutrino leakage scheme for the sake of computational efficiency in 3D. This has been replaced with multi-group, moment-based neutrino-transport schemes with algebraic closures \citep{martin21,kuroda20}
and other multi-group neutrino transport approximations \citep{powell23} in more recent studies. General-relativistic MHD codes were used in \citet{mosta14,mosta18,kuroda20} while Newtonian MHD codes with an approximate relativistic potential were used, e.g., in \citet{martin21,powell23}. Spatial resolution is another important numerical aspect and one needs to balance between the computational costs and the fidelity of the results. For example, \citet{martin21,reichert23,reichert24} adopted a coarse angular resolution of $\sim2.8^{\degree}$ and relatively low radial resolution of 210 zones for several 3D simulations. The smaller computational cost of low-resolution models allows a more systematic parametric survey for different progenitor models and combinations of rotation and magnetic fields. It is important, however, to complement these low-resolution models with higher resolution simulations as input for nucleosynthesis studies. Part of the motivation of this paper is to provide MR CCSN nucleosynthesis yields based on higher-resolution simulations.

In this paper, we perform nucleosynthesis calculations with the  \texttt{SkyNet} reaction network \citep{skynet} by 
post-processing the 3D non-MHD and MR-CCSN models of \citet{powell20,powell23}. The two 3D MR-CCSN models in \citet{powell23} span initial magnetic field strengths that differ by two orders of magnitudes with central fields of
$10^{10}\,\mathrm{G}$ and $10^{12}\,\mathrm{G}$, respectively.
While the non-MHD model exhibits a normal neutrino-driven explosion with a diagnostic explosion energy $E_{\rm exp}$ of $10^{51}$\,erg in $\sim1$\,s, in the MR models the shock is revived already $\sim100$\,ms after bounce and $E_{\rm exp}$ grows rapidly to $\sim 2.5\texttt{-}3\times10^{51}$\,erg in $\mathord{\sim}200$\,ms regardless of the different field strengths. The early and energetic explosion leads to neutron-rich ejecta with electron fractions down to $\mathord{\sim}0.25$ in the MR models. Accordingly our post-processing calculations show weak $r$-process nucleosynthesis up to the 2nd peak in the MR models and we observe insignificant differences for the different field strengths. The MR models also produce more radioactive $^{56}$Ni ($\sim$0.1-0.2\,$M_{\sun}$) than the non-MHD model ($\sim6.4\times10^{-2}\,M_{\sun}$) and more generally, normal CCSNe. The synthesized $^{56}$Ni mass in the MR models is in the ballpark of observed hypernovae \citep{bufano12,nomoto13,delia15}.

Our paper is organized as follows. In Section~\ref{sec:model} we summarize the CCSN progenitor and explosion models used in this work. In Section~\ref{sec:res} we present the methods used for post-processing calculations, including the extraction of tracers and \texttt{SkyNet} simulations. 
We then describe the nucleosynthetic conditions and detailed yields through the post-process calculations. We also discuss the potential astrophysical implications of our results. We give our conclusions in Section~\ref{sec:con}.

\section{Magnetorotational CCSN models}  \label{sec:model}
We base our nucleosynthesis calculations on two 3D MR-CCSN models from \citet{powell23} and one non-MHD CCSN model from \citet{powell20}. These were all simulated with the same progenitor model of \citet{aguilera18}, but with different magnetic field strengths imposed at the onset of core collapse. Below we briefly summarize essential features of the progenitor model and the results of CCSN hydrodynamic simulations to provide background for understanding the nucleosynthesis results. For a more comprehensive view of the models as well as simulation methods we refer the readers to the original papers.

\textit{Progenitor model: } The CCSN progenitor was evolved using the Modules for Experiments in Stellar Astrophysics (\textsc{MESA}, \citealt{mesa11,mesa13,mesa15,mesa18}) and has an initial mass of $39\,M_{\sun}$, initial metallicity of $\sim 1/50\,Z_{\sun}$, and initial rotational velocity of $600\,{\rm km\,s^{-1}}$. It is the most massive CCSN progenitor in the B Series models of \citet{aguilera18}, in which the diffusion constant due to rotational mixing is enhanced by a factor of 10. The model experiences chemically homogeneous evolution, develops a fast-rotating and massive carbon-oxygen core of $\mathord{\sim} 23\,M_{\sun}$, and loses the entire hydrogen and helium envelopes prior to collapse. Such an evolutionary path is considered to produce progenitor models for long GRB and Type Ic superluminous supernovae. 

At the onset of core collapse the model has a mass of $22.05\,M_{\sun}$ with a compactness parameter \citep{oconnor11} of $\xi_{2.5}=0.36$ and a core angular velocity of 0.54\,rad\,s$^{-1}$, which translates to a birth PNS period of 4.15\,ms assuming angular momentum conservation\footnote{This result assumes a proto-neutron star with a mass of 1.5\,$M_{\sun}$, a radius of 15\,km and a momentum of inertia of $2.27\times10^{45}$\,g\,cm$^2$.}. Magnetic fields are added to the progenitor model at the beginning of the core collapse simulations with a maximum strength of $10^{10}$\,G (model m39\_B10) and $10^{12}$\,G (model m39\_B12) at the stellar center for both the poloidal and toroidal components. A dipolar magnetic field is assumed and generated from the vector potential $\Vec{A}$ given by \citep{suwa07}
\begin{equation}
    (A^r,A^\theta,A^\phi) = \frac{B_0r_0^3r}{2(r^3+r_0^3)}\big(\cos\theta,0,\sin\theta),
\end{equation}
where $B_0$ is the maximum and central field strength and $r_0$ is set to 10$^3$\,km.
For the non-MHD model (m39\_B0) no magnetic field is imposed. This model is used as a reference neutrino-driven CCSN model.

\textit{CCSN models}: The CCSN simulations were performed using the \textsc{CoCoNuT-FMT} code \citep{dimmelmeier02,muller15,powell19} with the MHD solver described in \citet{muller20}. Neutrino transport is solved using the 3-flavor ($\nu_e$, $\bar{\nu}_e$ and $\nu_x=\{\nu_\mu,\bar{\nu}_\mu,\nu_\tau,\bar{\nu}_\tau\}$) multi-energy-group one-moment scheme (FMT) of \citet{muller15}. Neutrino energy is sampled with 21 logarithmically distributed bins from 4\,MeV to 240\,MeV.  The 3D computational grid has $550\times128\times256$ zones in radius, latitude, and longitude, reaching $10^5$\,km
in radius, and covers the full 4$\pi$ solid angle. This corresponds to an angular resolution of $\mathord{\sim}1.4^{\degree}$. The finest resolution for the radial grid is 192\,m at the PNS center, and the grid maintains $\Delta r/r\sim0.015$ from the outer layers of PNS to a few hundred km. For comparison, the 3D MHD simulations in \citet{martin21} used a two-moment scheme (M1) for the neutrino transport with 10 energy bins, an angular resolution of $\mathord{\sim}2.8^{\degree}$ and 300 radial zones out to an outer grid boundary at $5\times 10^{6}\,\mathrm{km}$.

The simulations were performed in 2D during the collapse phase and mapped to 3D shortly after the core bounce with small perturbations imposed to trigger non-axisymmetric instabilities. The non-MHD model m39\_B0 used the general-relativistic version of \textsc{CoCoNuT} while the MR-CCSN model used its Newtonian MHD version \citep{muller20} with the effective relativistic gravitational potential (Case Arot) of \citet{muller08}. The 3D simulations stop at $\sim1\,$s, $\sim300$\,ms and $\sim680$\,ms after core bounce for m39\_B0, m39\_B10 and m39\_B12, respectively.

\begin{figure}
    \centering
    \includegraphics[width=0.47\textwidth]{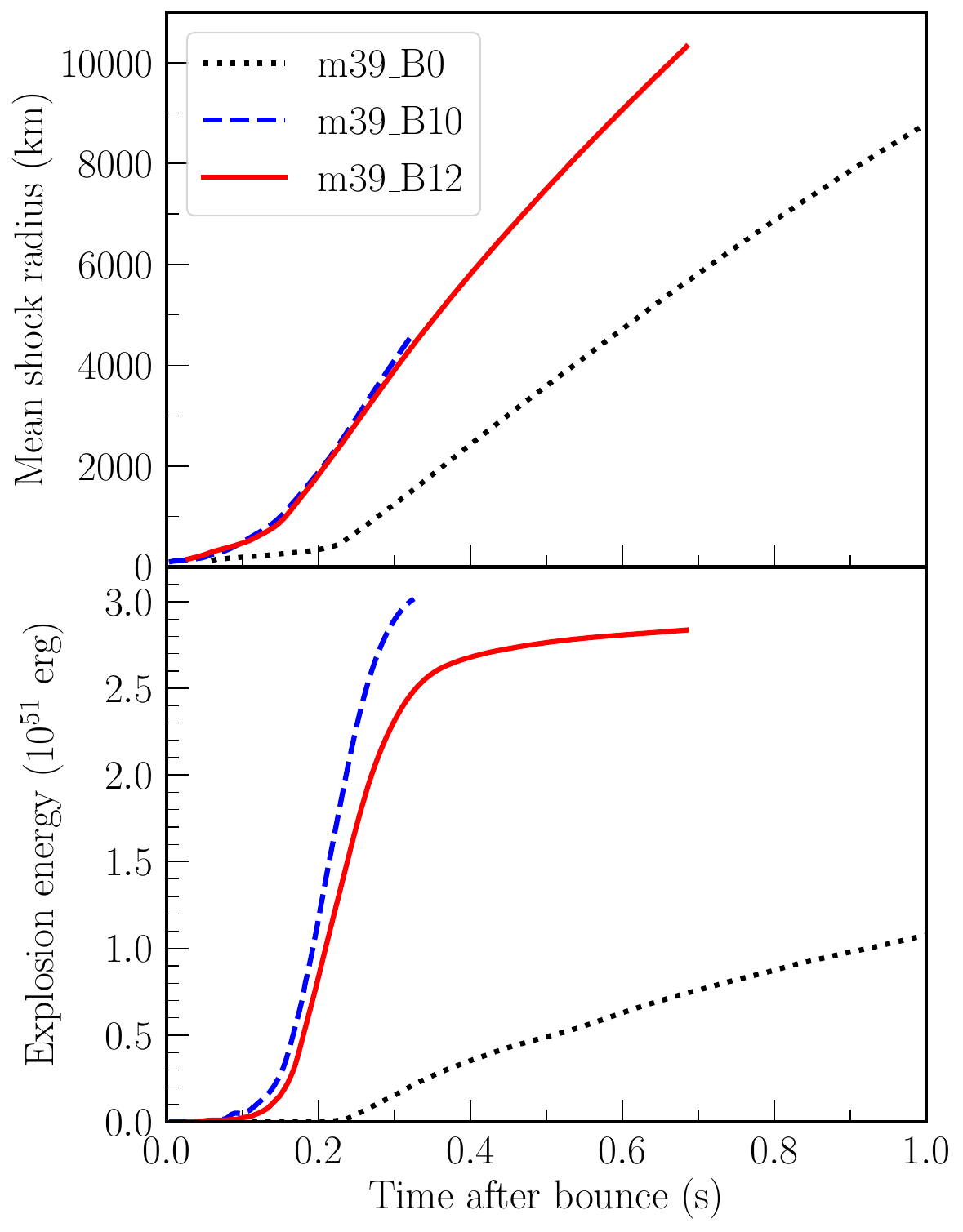}
    \caption{Evolution of the mean shock radius (upper panel) and the diagnostic explosion energy (lower panel) as a function of  time after bounce in our  3D supernova models. Model m39\_B0 does not contain magnetic fields, while the models m39\_B10 and m39\_B12 include magnetic fields. }
    \label{fig:exp}
\end{figure}

The explosion dynamics for these CCSN models are illustrated in Figure~\ref{fig:exp}, which shows the time evolution of the mean shock radius and the diagnostic explosion energy ($E_{\rm exp}$). Here diagnostic explosion energy is defined as the energy of material with positive radial velocity and total energy (summing over kinetic, internal, gravitational, and magnetic energies) in the simulated domain. This is different from the final explosion energy because there is still energy input from neutrino heating and nuclear recombination. Also, energy will be drained to unbind the overburdened envelope as the shock propagates outwards. Long-term simulations are needed to account for these further effects and determine the final explosion energy (see, e.g., \citealt{chan20,bollig21}). The MR models undergo shock revival at $\mathord{\sim}100$\,ms after core bounce with positive $E_{\rm exp}$ emerging, and this is $\mathord{\sim}100$\,ms earlier than that in the non-MHD model. $E_{\rm exp}$ in the MR models grows rapidly for about 200\,ms and reaches $\sim3.0\times10^{51}$\,erg in m39\_B10 and $\sim2.5\times10^{51}$\,erg in m39\_B12 at $\sim300$\,ms after core bounce. In the meantime, the total angular momentum of the PNS drops dramatically by an order of magnitude. This signifies the efficient tapping of the PNS rotational energy reservoir by the strong magnetic fields to power the explosion. The rise of $E_{\rm exp}$ turns much shallower afterwards, with an increment of $\sim0.3\times10^{51}$\,erg for the remaining $\sim$380\,ms in the model m39\_B12. A slow, sustained rise even after spindown of the the PNS is expected because there will still be low-level energy input into the explosion by neutrino heating, and possibly from magnetic fields that tap accretion power or the slow late-time rotation of the PNS. However, because the accretion rate in the magnetic models is low at this stage, this late-time rise is slower than in the non-magnetic model, which maintains a higher accretion rate and hence stronger neutrino heating.
An interesting observation is that the two MR models show very similar explosion dynamics though their initial magnetic field strengths differ by two orders of magnitudes. As we shall see, this has a major impact in the form of close similarity between their nucleosynthesis yields. In the non-MHD model, $E_{\exp}$ grows slowly as the explosion sets in and reaches $\sim10^{51}$\,erg at $\sim1.0$\,s after core bounce, which is still faster than the typical 3D non-rotating CCSN simulations of less massive progenitors (e.g., \citealt{burrows20,bollig21}). The growth of $E_{\exp}$ has not yet saturated in this model. At the end of simulations, the PNS baryonic mass is $\sim$2.03, 1.71 and 1.73$\,M_{\sun}$ in m39\_B0, m39\_B10 and m39\_B12, respectively. For computing mass fractions and production factors in the ejecta, we assume that the total ejecta mass is $\sim20\,M_{\sun}$ including the inner ejecta as simulated plus the unshocked outer envelope. Fallback accretion onto the PNS is
a potential caveat here, as this may reduce the final ejecta mass. This effect would need to be determined by long-term simulations \citep{chan20,burrows23}.

\begin{figure*}
    \centering
    \includegraphics[width=0.95\textwidth]{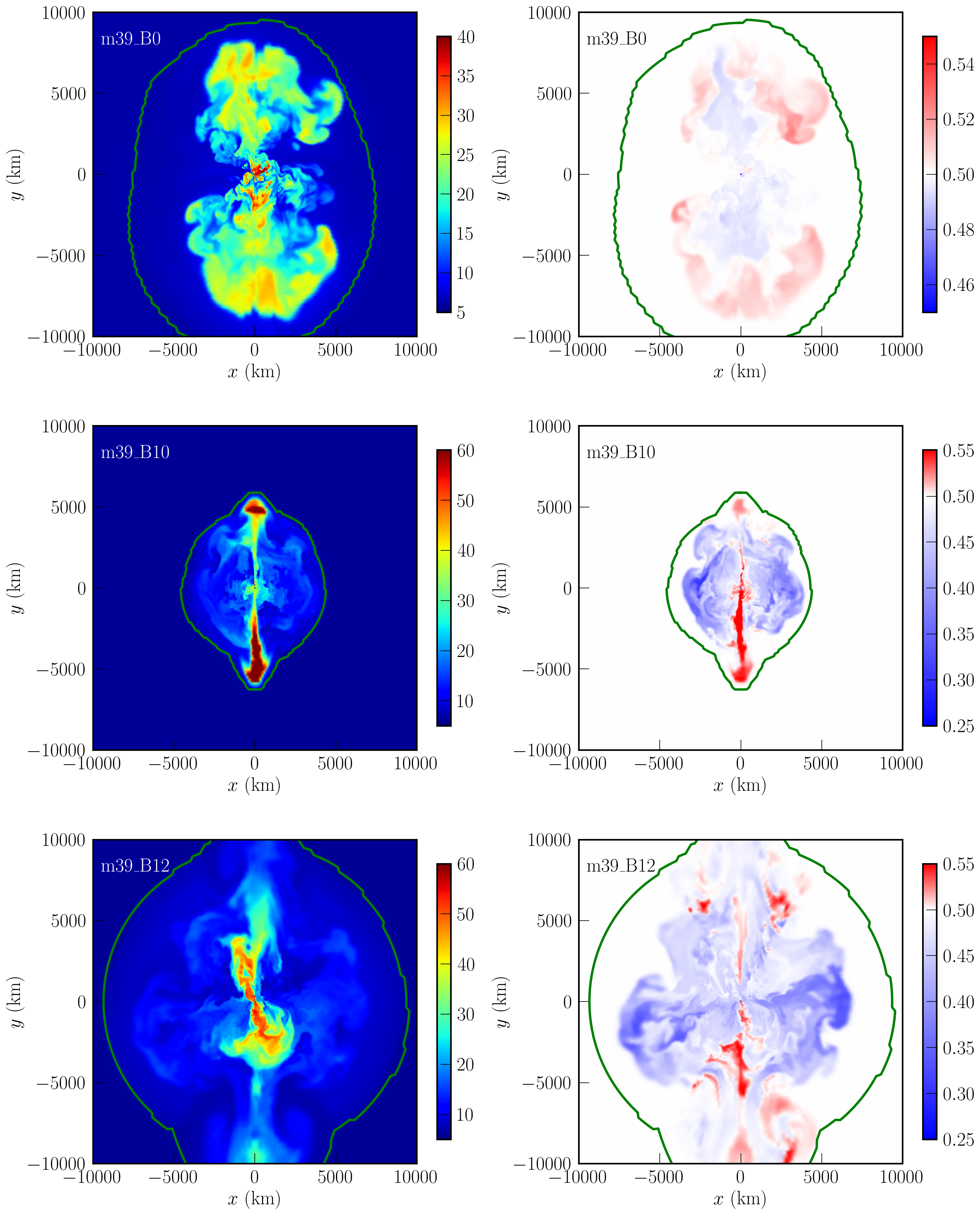}
    \caption{Meridional slices showing specific entropy (left column, in units of $k_{\rm B}$ per nucleon) and electron fraction (right column) for the models near the end of simulations, i.e. m39\_B0 at 1\,s (top row), m39\_B10 at 0.3\,s (middle row) and m39\_B12 at 0.68\,s (bottom row). Note that the jets in the model m39\_B12 are less collimated and more off-axis compared to the model m39\_B10 due to the operation of non-axisymmetric instabilities during the longer simulated time. The green curve in each panel marks the corresponding shock surface.}
    \label{fig:slice}
\end{figure*}

To view the CCSN ejecta configuration in the adopted 3D models, we render the specific entropy and electron fraction $Y_\mathrm{e}$ on a meridional plane at the end of simulations in Figure~\ref{fig:slice}. All three models exhibit bipolar features along the rotational axis to some degree. In the MR models the supernova shock and the distribution of the bulk of the ejecta is relatively spherical while bipolar jets are launched $\sim50$-$100$\,ms after shock revival. Run until $\sim300$\,ms after core bounce, the model m39\_B10 has obvious collimated bipolar jets, though the jets remain non-relativistic. In model m39\_B12, which has been run for $\sim$400\,ms longer, the jets are less collimated and off-axis due to non-axisymmetric kink instabilities. The jets in m39\_B10 may also experience such non-axisymmetric instabilities and evolve similarly as m39\_B12. The ejecta for the bulk shock are mostly mildly to moderately neutron-rich with $Y_\mathrm{e}$ down to $\sim$0.25, while the jets contain proton-rich ejecta with $Y_\mathrm{e}>0.5$ as clearly seen in m39\_B10. Similar proton-rich jets are found in the 3D model `O' in \citet{reichert23}. In comparison, the explosion in m39\_B0 is more asymmetric and bipolar in terms of bulk ejecta distribution and shock geometry and $Y_\mathrm{e}$ of the ejecta lies in the interval of $[0.45,0.55]$. 

It should be emphasized that although jets are present in the MR models, they are not dominant for driving the explosion. The shock revival and rapid growth of $E_{\rm exp}$ take place before the jet formation. Also, the jets carry at most $\sim10\%$ of $E_{\rm exp}$ and their composition is proton-rich. The neutron-rich ejecta are related to the early shock revival and the rapid growth of $E_{\rm exp}$. Similar prompt ejection of neutrino-rich matter is also found the 3D MR models of \citet{reichert23}. We note that an early explosion with fast shock expansion is more favorable for neutron-rich ejecta in the neutrino-driven explosion scenario. This is true for the electron-capture supernova and low-mass CCSN explosions (e.g., $Y_\mathrm{e}\sim0.4$, \citealt{wanajo18}). For more massive progenitors, \citet{wanajo18} found that in their s27 model the ejecta $Y_\mathrm{e}$ extends down to 0.40 (otherwise $\gtrsim$0.45 for other massive progenitors) as it also explodes early because of prominent standing accretion shock instability \citep{muller12}. 

\section{Nucleosynthetic yields} \label{sec:res}

\subsection{Nucleosynthetic conditions} \label{ssec:condition}
Before discussing the yields, it is useful to look at the nucleosynthesis conditions in the CCSN ejecta. We sample the fluid elements with tracer particles to get the evolution history of the ejecta. Tracer trajectories, i.e. density and temperature as a function of time are extracted from the recorded snapshots of simulations with a cadence of $\sim1$\,ms. We sample the tracers in the final snapshot and then use the backward integration method as described in \citet{wanajo18,sieverding20} to obtain trajectories. \citet{sieverding23} showed that this backward algorithm has advantages over the forward approach and the results agree reasonably well with traditional on-the-fly tracer trajectory integration in which particle tracers co-move with the fluids during the hydrodynamic evolution \citep{nagataki97,harris17}. We sample the final snapshot with $30\times24\times40$ bins in the $r$, $\theta$ and $\phi$ directions, covering logarithmically from $\sim450$\,km to the maximum shock radius and covering $4\pi$ in solid angle. In particular, for the $\theta$ direction we use equal $\Delta(\cos\theta)$ for the sampling to give each tracer the same volume. The total number of tracers used is thus 28800 and each tracer particle has a mass ranging from $\sim10^{-8}\,M_{\sun}$ to $\sim10^{-3}\,M_{\sun}$. We also justify our results against the tracer resolution by doubling the number of bins in all directions which results in negligible difference in the final yield pattern.

For defining the ejecta region, we use a conservative criterion that a tracer particle is ejected only if its final total energy (summing over internal, kinetic, gravitational and magnetic) and radial velocity are both positive. As we are most interested in heavy-element production, we only post-process the tracers with a peak temperature $T_{\rm p}\ge3.5$\,GK below which the synthesis of Fe-group and trans-Fe elements is negligible. The resulting numbers (total mass) of the ejected tracers are 12138 ($0.124\,M_{\sun}$), 21826 ($0.518\,M_{\sun}$) and 12253 ($0.606\,M_{\sun}$) for models m39\_B0, m39\_B10 and m39\_B12, respectively.

We use two post-processing approaches for the tracers depending their $T_{\rm p}$ (see also \citealt{reichert23,wang23}). For $T_{\rm p}\ge T_{\rm p,th}=7\,$GK, the calculation starts from the nuclear statistical equilibrium (NSE) composition when the tracer's temperature drops below 7\,GK. We take $Y_\mathrm{e}$ at that moment to get the initial NSE composition. For tracers with $T_{\rm}<7$\,GK, the calculation starts from the core bounce with the initial composition of the progenitor model (and thus with initial $Y_\mathrm{e}=0.5$ in the relevant region). These low-$T_{\rm p}$ tracers have collapsed to a minimum radius greater than $\sim1000$\,km and their $Y_\mathrm{e}$ is only minutely affected by neutrino interactions. We have checked that the choice of $T_{\rm p,th}$ do not affect our major results if varied from 7\,GK to 8\,GK.

\begin{figure}
    \centering
    \includegraphics[width=0.47\textwidth]{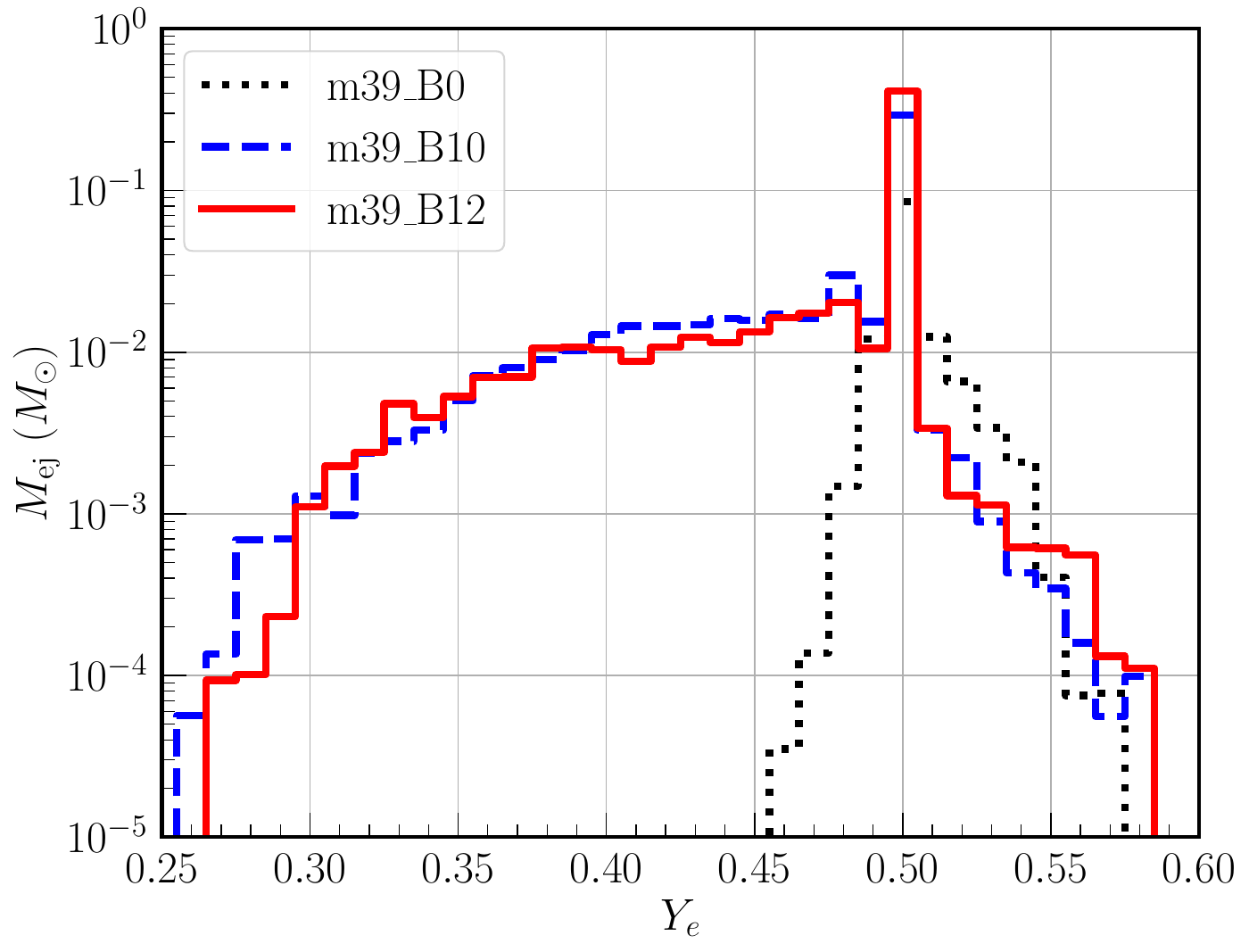}
    \caption{Distribution of ejecta mass $M_{\rm ej}$ per bin as a function of the electron fraction $Y_\mathrm{e}$ with a bin width of 0.01. }
    \label{fig:mvsye}
\end{figure}

$Y_\mathrm{e}$ is the dominant factor that determines the heavy-element nucleosynthesis for tracers reaching NSE. The mass distribution of the chosen ejected tracers as a function of $Y_\mathrm{e}$ is shown in Figure~\ref{fig:mvsye}. In accord with the slice plots in Figure~\ref{fig:slice}, the ejecta in the two MR models have a long neutron-rich tail with $Y_\mathrm{e}$ extending down to $\sim0.25$. The mass distribution has a shallow slope against $Y_\mathrm{e}$ in the interval from $\sim0.35$ to $\sim0.5$. Due to their very similar explosion dynamics (c.f.\ Figure~\ref{fig:exp}), the $Y_\mathrm{e}$ distributions in the 2 MR models
resemble each other closely\footnote{We note that in the Figure 9 of \cite{powell23}, the mass distribution is normalized to the total ejecta mass. This results in the smaller value for the neutron-rich part at 680\,ms. In fact, the absolute mass distribution of neutron-rich ejecta is almost identical at 300\,ms and 680\,ms.}. In comparison, the non-MHD model shows a distribution for typical neutrino-driven CCSN ejecta with $Y_\mathrm{e}$ more or less centered around 0.5, with a minimum $Y_\mathrm{e}\sim0.45$ and a more extended distribution on the proton-rich side (e.g., \citealt{wanajo18,wang23}).

We note that all the 3D CCSN models here show a steep drop of the ejecta distribution towards high $Y_\mathrm{e}$ above 0.5, which tends to be shallower in CCSN simulations with more accurate neutrino transport \citep{wanajo18,wang23}. This likely stems from the difference in the average energies of the electron-flavor neutrinos ($\Delta\varepsilon=\langle \varepsilon_{\bar{\nu}_e} \rangle-\langle \varepsilon_{{\nu}_e} \rangle$, see \citealt{sieverding20}). $\Delta\varepsilon$ is $\sim2$\,MeV in accurate neutrino transport schemes while it is $\sim3-4$\,MeV during the explosion in the simplified FMT scheme \citep{muller15} which is used in our 3D CCSN models. Since the absorption of $\nu_e$ lifts $Y_\mathrm{e}$, this raises
the asymptotic equilibrium $Y_\mathrm{e}$ under neutrino exposure, which is determined by
the luminosities $L_{{\nu}_\mathrm{e}}$ and
$L_{\bar{\nu}_\mathrm{e}}$ and
mean energies $\varepsilon_{{\nu}_\mathrm{e}}$ and
$\varepsilon_{\bar{\nu}_\mathrm{e}}$
of electron neutrinos and antineutrinos \citep{qian_96},
\begin{equation}
Y_e \approx \left[1+\frac{L_{\bar{\nu}_e} (\varepsilon_{\bar{\nu}_e}-2 \Delta)}
{L_{{\nu}_e} (\varepsilon_{{\nu}_e}+2 \Delta)}\right]^{-1}.
\label{eq:ye_equilibrium}
\end{equation}
However, this may not significantly change $Y_\mathrm{e}$ of the bulk neutron-rich ejecta in the MR models, which result from freeze-out of the $Y_\mathrm{e}$ well below its asymptotic equilibrium value anyway.
Another uncertainty around $Y_\mathrm{e}$ concerns the approximate treatment of general relativistic effects in our simulations by means of an effective potential in the MHD simulations (as opposed to the general relativistic simulation without magnetic fields).
For core-collapse supernova simulations without magnetic fields, \citet{muller12} demonstrated that simulations with an effective potential approximate the neutrino luminosities and mean energies in relativistic simulations fairly well. However, they found electron neutrino and antineutrino luminosities to be slightly higher in the general relativistic case due to subtle changes in proto-neutron star structure. Higher luminosities could push $Y_\mathrm{e}$ towards the asymptotic equilibrium value
(Equation~\ref{eq:ye_equilibrium}) more quickly. There may also be a small shift in the equilibrium $Y_\mathrm{e}$ itself, but this likely has a very minor effect on the $Y_\mathrm{e}$-distribution. The bottom line of these uncertainties is that the $Y_\mathrm{e}$-distribution in the two magnetorotational cases may be a little less neutron-rich than predicted by the simulations, but the presence of much more neutron-rich  ejecta than in the neutrino-driven case is expected to be robust.

\begin{figure*}
    \centering
    \includegraphics[width=0.97\textwidth]{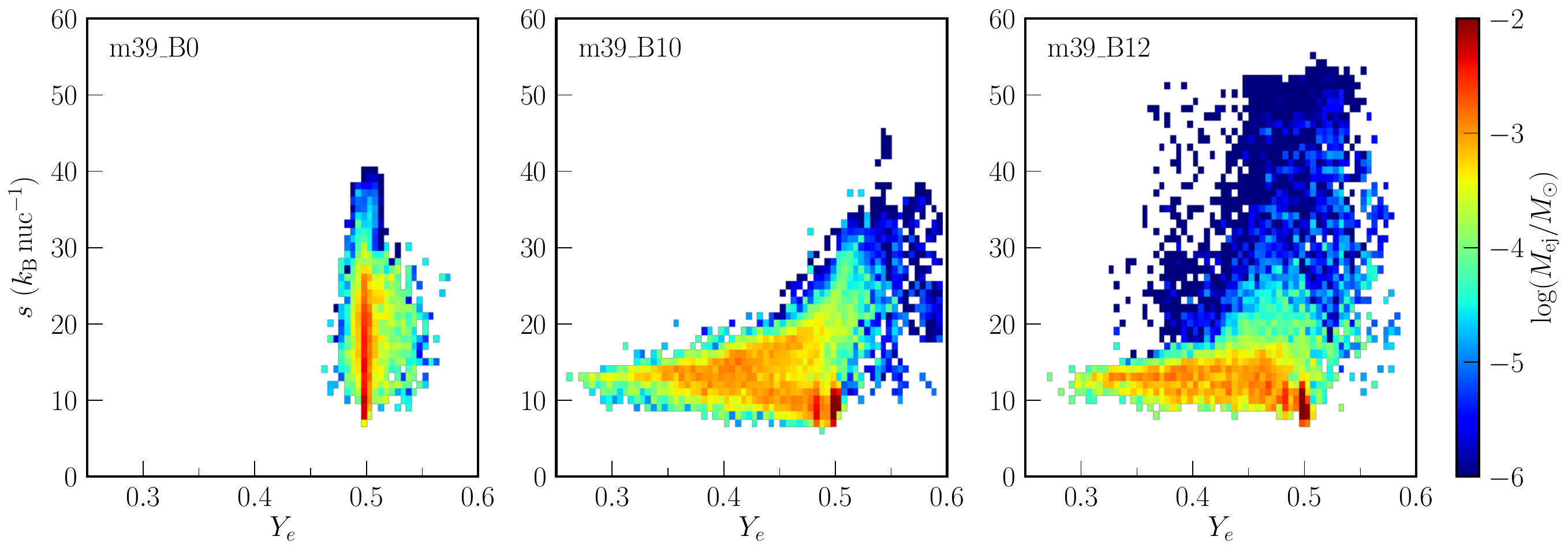}
    \caption{Distribution of the ejecta mass $M_{\rm ej}$ as a function of the electron fraction $Y_\mathrm{e}$ and the asymptotic entropy per nucleon $s$. Note that the model m39\_B10, the high-entropy jets (c.f. the middle row in Figure~\ref{fig:slice}) are not well sampled by the tracers due to the small opening angle ($\sim10^{\degree}$), so ejecta with $s\gtrsim40\,{k_{\rm B}\,{\rm nuc}^{-1}}$ are lacking. }
    \label{fig:mvsyes}
\end{figure*}

Figure~\ref{fig:mvsyes} further shows the same mass distributions, but as a function of $Y_\mathrm{e}$ and the asymptotic entropy per nucleon ($s$). The low $Y_\mathrm{e}$ components of the ejecta in the MR models correspond to low values of entropy, $\sim10-20\,k_{\rm B}\,{\rm nuc}^{-1}$. This reflects that they are associated with the early onset of explosion and less affected by neutrino heating. The proton-rich side of ejecta can reach high values of entropy $\sim40-50\,k_{\rm B}\,{\rm nuc}^{-1}$, because much of these components are associated with the jets. A large spread in $Y_\mathrm{e}$ is seen for the high-entropy components in m39\_B12, as the jets are mixed with the bulk ejecta in later times due to non-axisymmetric instabilities. A caveat is that this large spread may be partly due to numerical sampling and numerical diffusion.

Finally, the jets in model m39\_B10 exhibit high-entropy ($s\gtrsim40\,{k_{\rm B}\,{\rm nuc}^{-1}}$) and proton-rich ($Y_\mathrm{e}>0.5$) components which are not seen in the $Y_\mathrm{e}-s$ distribution of tracers in Figure~\ref{fig:mvsyes}. This is due to the small opening angle ($\sim10^{\degree}$) of the jets at the early times (recall that this model was simulated for a shorter duration after bounce) which is not well sampled when binned with $\Delta(\cos\theta)$. To quantify the impact of the jets on our results, we additionally sample their volume with a finer resolution in $\theta$. We use 6 bins in the $\theta$ direction to sample the intervals of $[0^{\degree},15^{\degree}]$ and $[165^{\degree},180^{\degree}]$. These volumes contribute $\sim0.014\,M_{\sun}$ mass which is $\sim3\%$ of the total mass of the inner ejecta. Because this region mainly contributes high-entropy and proton-rich material, it will not be relevant for $r$-process nucleosynthesis.

\subsection{Detailed nucleosynthetic yields}
\label{sec:detailed_nucleo}
We use the \texttt{SkyNet} code \citep{skynet} to perform the post-processing nucleosynthesis calculations for the above extracted tracer trajectories. To explore the possibility of $r$-process production in MR CCSNe, we use the reaction network with 7836 isotopes up to $^{337}_{112}$Cn. Settings of the reaction network are following the setup from `$r$-process.py' in the examples of \texttt{SkyNet} with reaction rates from JINA REACLIB database \citep{reaclib}. We do not include neutrino interactions in our post-processing calculations. Further neutrino absorption can raise $Y_e$ of the inner ejecta in general (see a detailed discussion in \citealt{sieverding20}). Intense neutrino fluxes can also affect heavy-element nucleosynthesis through $\nu$-process \citep{woosley1990} and $\nu p$-process \citep{frohlich2006} at low temperatures. We leave these effects on the nucleosynthesis of MR CCSNe for future studies.

Because the tracers still have a high temperature ($\gtrsim2$\,GK) at the end of hydrodynamic simulations, we extrapolate the tracer trajectories to $5\times10^7$\,s according to the following equations \citep{arcones07,ning07}
\begin{equation}
    \begin{aligned}
        \rho(t) & = \rho(t_{\rm e})[1+(t-t_{\rm e})/\tau]^{-2}, \\
        T(t) & = \rho(T_{\rm e})[1+(t-t_{\rm e})\tau]^{-2/3}, 
    \end{aligned}
\end{equation}
where $t_{\rm e}$ is the time when the (magneto-)hydrodynamic simulations terminate, and we fit the last $\sim6$\,ms of the trajectories to get the expansion timescale $\tau$. 

As mentioned in Section~\ref{ssec:condition}, for tracers with $T_{\rm p}\ge$7\,GK we start the calculation from NSE with $Y_\mathrm{e}$ at $T=7$\,GK, and for tracers with 3.5\,GK$\le T_{\rm p}<$7\,GK we start the calculation from the time of bounce and use the initial abundance from the progenitor model, which is evolved with the 21-isotope network of \textsc{MESA}. Note that this network includes mainly $\alpha$-isotopes and a few neutron-rich isotopes ($^{56}$Cr, $^{54}$Fe, $^{56}$Fe) for the low-$Y_\mathrm{e}$ Fe core. Therefore, the low-$T_{\rm p}$ tracers mostly coming from the Si-shell have an initial composition of O, Si, Si and Ca. This may overestimates the production of radioactive $^{56}$Ni and underestimates synthesis of stable neutron-rich Fe-group isotopes like $^{58}$Ni and so on. Future study should consider evolving the progenitor with a larger network.
%or perform post-processing calculation for the pre-collapse stage. 

\begin{figure*}
    \centering
    \includegraphics[width=0.97\textwidth]{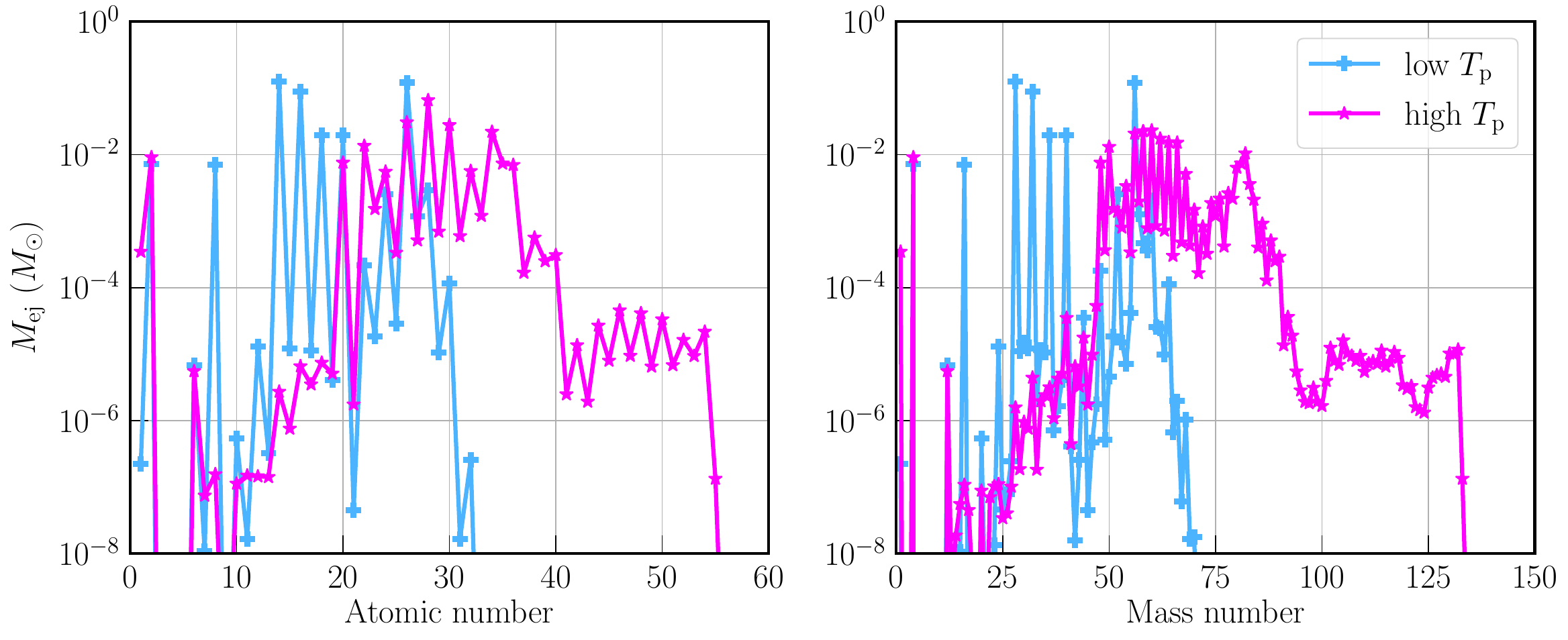}
    \caption{Comparison of the ejecta mass ($M_{\rm ej}$) distribution as a function of the atomic number (left panel) and the mass number (right panel) between the tracers with a high and low peak temperature ($T_{\rm p}$) in the model m39\_B12. The separating $T_{\rm p}$ is 7\,GK for the high- and low-$T_{\rm p}$ tracers.}
    \label{fig:B12_Tp}
\end{figure*}

\begin{figure*}
    \centering
    \includegraphics[width=0.97\textwidth]{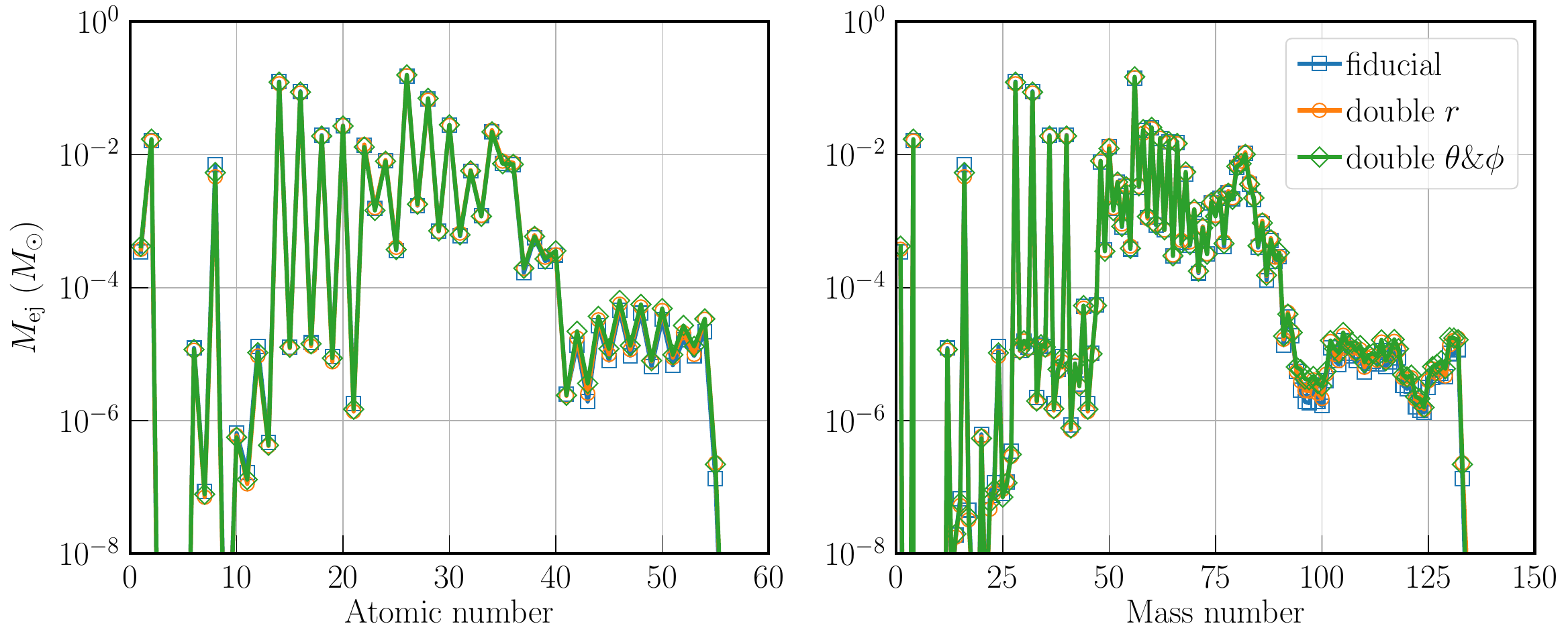}
    \caption{A convergence test of the resolution used in the tracer sampling for the model m39\_B12. The `fiducial' is the default resolution used in all our models. The `double $r$' is doubling the resolution in $r$-direction, and the `double $\theta\&\phi$' is doubling the resolution in both $\theta$ and $\phi$ directions. }
    \label{fig:B12_res}
\end{figure*}

For model m39\_B12, the high- and low-$T_{\rm p}$ tracers contribute 0.209\,$M_{\sun}$ and 0.397\,$M_{\sun}$, respectively. We compare the absolute yield mass distribution after radioactive decays
as a function of atomic number $Z$ and mass number $A$ between the high- and low-$T_{\rm p}$ tracers in Figure~\ref{fig:B12_Tp}. As expected, the low-$T_{\rm p}$ tracers mainly produce $\alpha$-elements from $^{28}$Si to $^{40}$Ca and $^{56}$Ni (here decayed to $^{56}$Fe). The high-$T_{\rm p}$ tracers mainly produce Fe-group and trans-Fe elements up to atomic number 36 with $\sim10^{-3}-10^{-2}\,M_{\sun}$, with robust synthesis of the first $r$-process peak ($A\sim80$). They can also synthesis isotopes until the second $r$-process peak ($A\sim 130$) in low quantities with $\sim10^{-5}\,M_{\sun}$. Lastly, the high- and low-$T_{\rm p}$ tracers produce $\sim0.021\,M_{\sun}$ and $\sim0.120\,M_{\sun}$ of $^{56}$Ni, respectively. In combination the $^{56}$Ni yields in model m39\_B12 is $\sim0.141\,M_{\sun}$. This value is a factor of $4\texttt{-}5$ larger than that estimated by the flashing scheme \citep{rampp02} used in the \textsc{CoCoNuT-FMT} hydrodynamic simulation. Figure~\ref{fig:B12_res} shows the results of a convergence test of the resolution used in the tracer sampling for the model m39\_B12. We have performed two additional runs with higher resolutions than that of the fiducial run. One is doubling the resolution in $r$-direction (double $r$), and the other is doubling the resolution in both $\theta$ and $\phi$ directions (`double $\theta\&\phi$'). We found excellent agreements among these post-process nucleosynthesis calculations.

\begin{figure*}
    \centering
    \includegraphics[width=0.97\textwidth]{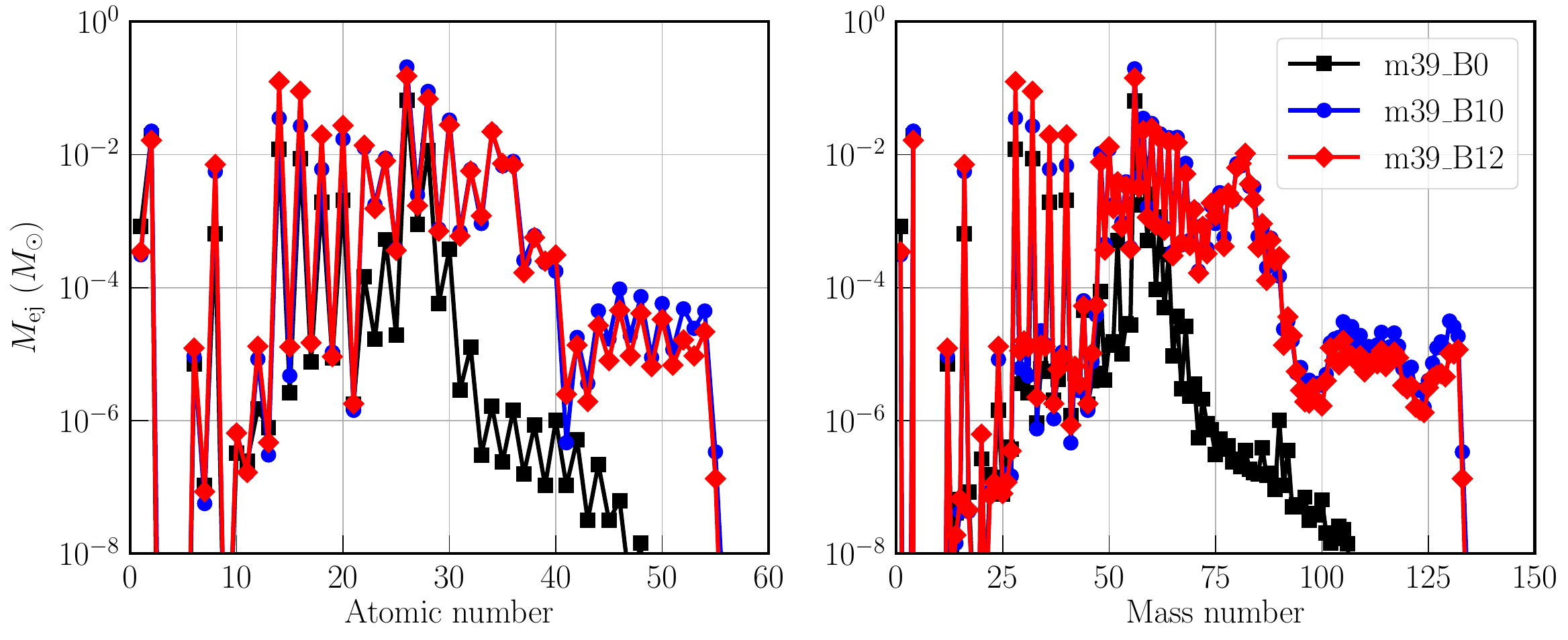}
    \caption{Comparison of the ejecta mass ($M_{\rm ej}$) distribution as a function of atomic number (left panel) and mass number (right panel) for the three 3D supernova models.}
    \label{fig:MvsZA}
\end{figure*}

In Figure~\ref{fig:MvsZA} we compare the absolute yields of the innermost ejecta among our three CCSN models. The two MR models show quite similar patterns with subtle differences, mostly noticeable for the $\alpha$ elements. Fair agreement is found for the distribution of $Z\sim20-40$ and $A\sim40-100$. For heavier elements ($Z>40$ and $A>100$), the yields in m39\_B10 are a factor of $\sim2$ larger that those in m39\_B12. The heaviest non-negligible isotope is $^{133}_{~55}$Cs in the MR models.  Given the similarity in the yields beyond Fe between the two MR models, we expect that the results of heavy elemental production would not change substantially if model m39\_B10 was simulated for longer. On the other hand, the non-MHD model can only synthesize isotopes up to $Z\sim48$ and $A\sim105$, similar to the yield patterns of normal CCSNe with massive progenitors ($\gtrsim10\,M_{\sun}$, e.g., \citealt{wanajo18,wang23}). The $^{56}$Ni yields are $0.064\,M_{\sun}$, $0.194\,M_{\sun}$ and $0.141\,M_{\sun}$ in m39\_B0, m39\_B10 and m39\_B12, respectively. The stronger production of $^{56}$Ni and less production of $\alpha$ elements in m39\_B10 is due to its larger explosion energy and thus bigger mass of ejecta with a higher $T_p$ compared to m39\_B12. The $^{56}$Ni mass and explosion energy in the MR models suggest that they may represent some observed events on the low end of the hypernova population \citep{bufano12,nomoto13,delia15}, though detailed radiative transfer modelling is needed to justify the applicability to observed transients.

Because magnetically collimated jets are not the main driver of the MR explosions considered in this study, and because the jets launched after the bulk explosion are made up of proton-rich components, we found no elements synthesized beyond the second $r$-process peak. This is different from the simulations with parametrized jets in \citet{nishimura15,grimmett21} and early 3D MHD simulations resulting in jet-driven explosions \citep{winteler12,mosta18,halevi18}, in which the ejecta have neutron-rich components with $Y_\mathrm{e}$ down to $\sim$0.1, allowing $r$-process nucleosynthesis up to the third peak. Our results closely resemble the recent 3D models of \citet{reichert23}. More specifically, the yield patterns of the MR models here fall between their `O' and `P' models, where `O' has a moderate magnetization with poloidal (toroidal) strength of $1.7\times10^{10}$ ($1.7\times10^{11}$) G and `P' has a 3 times stronger poloidal component than `O'. This suggests that MR explosions without strong collimated jets can be a viable and robust source for weak $r$-process. Producing heavier elements 
%in extremely metal-poor stars 
would require MR explosions with either stronger magnetic fields and specific field configuration \citep{reichert24}, or other explosion scenarios (e.g., collapsars \citealt{siegel19}).

\subsection{Astrophysical implications}
The imprint of nucleosynthesis for any individual supernova type can be seen mainly in two ways. One is to compare the yield distribution in theoretical models with the observed abundance pattern of individual stars. This is particularly useful for (extremely) metal-poor stars as their formation site may have been polluted by just a single supernova ever occurred there (e.g., \citealt{keller14,yong21,xing23}). Another way is to consider the over-produced isotopes relative to the Sun to examine the resulting contribution of a supernova type to metal enrichment in galactic chemical evolution (GCE). One can roughly constrain the fraction of a particular CCSN type relative to the whole CCSN population by looking at the most over-produced isotopes \citep{wanajo11}. Below we explore the latter possibility for our MR-CCSN models.

\begin{figure*}
    \centering
    \includegraphics[width=0.97\textwidth]{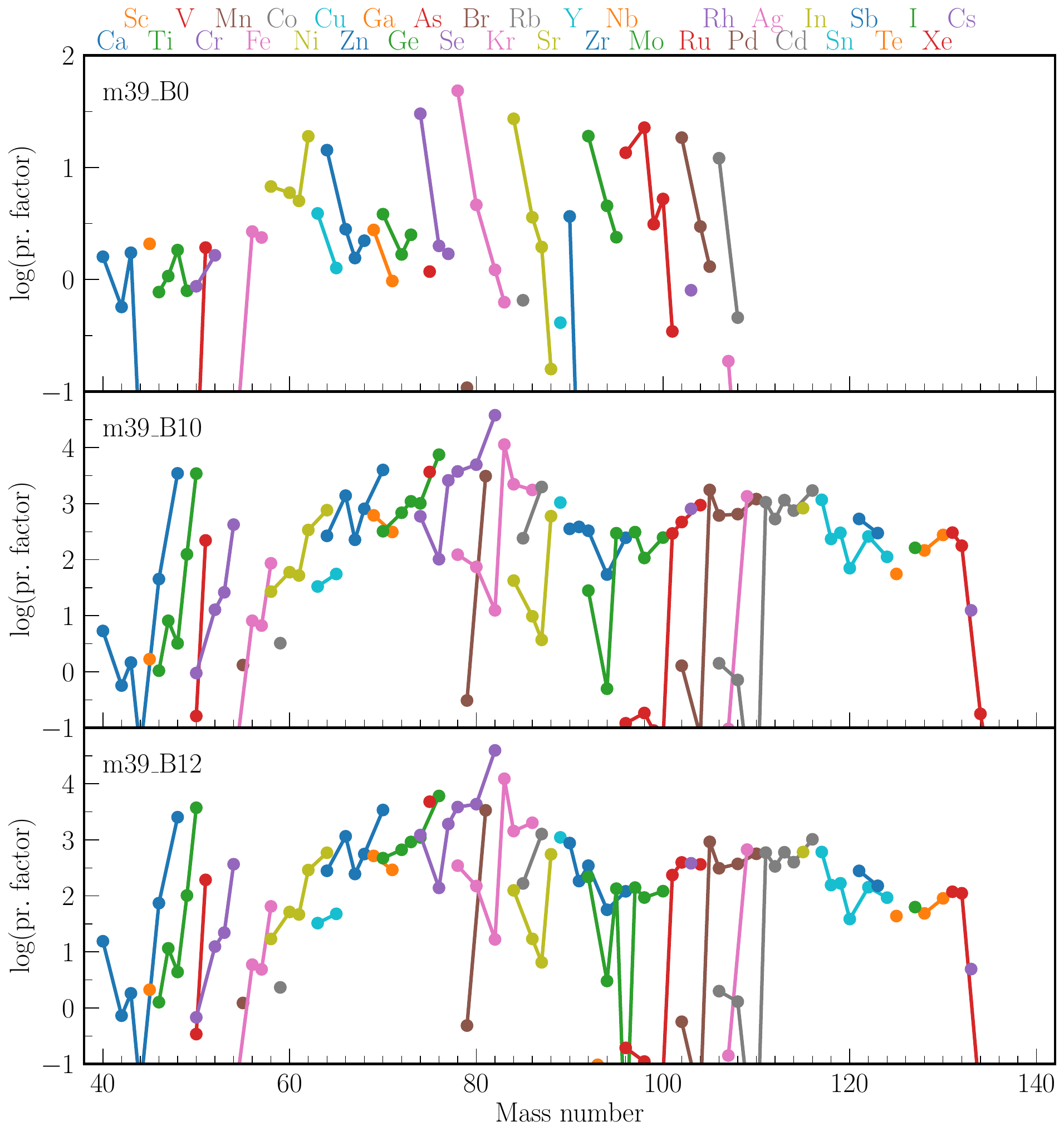}
    \caption{Isotopic production factors as a function of mass number from Ca to Cs for models
    m30\_B0, m30\_B10 and m30\_B12(top to bottom).}
    \label{fig:pf_A}
\end{figure*}

To estimate the potential contribution to GCE, we calculate the production factor of each isotope $i$ by
\begin{equation}
    {\rm pr.~factor} = X_i / X_{i,\sun},
\end{equation}
where $X_i$ and $X_{i,\sun}$ are the mass fractions in the simulated CCSN models and the Sun (taken from \citealp{lodders09}), respectively. For the models, we compute $X_i$ as the quotient between the synthesized isotopic yield in the innermost ejecta and the total ejecta mass $\sim20\,M_{\sun}$, i.e. including the unshocked outer material by the end of the simulation. This approximation is justifiable for the heavy elements including the Fe-group and trans-Fe regime because no further production of these elements occurs during shock propagation through the envelope.  The isotopic production factors are shown in Figure~\ref{fig:pf_A} for all the models. Explicitly, these isotopic production factors ignore the contribution of outer unshocked material which will have a peak temperature below $\sim3\,$\,GK. However, fallback of Fe-group and trans-Fe material is a source of uncertainty and might decrease the predicted production factors.

The largest production factor is $\sim4\times10^4$ for $^{82}_{34}$Se in both MR models and $\sim50$ for $^{78}_{32}$Kr in the non-MHD model. There is a trend that the MR models favor the production of more neutron-rich isotopes due to the overall low $Y_\mathrm{e}$ of their innermost ejecta. Elements beyond Zn and 
up to Xe generally have production factors larger than $\sim100$ in the MR models. This implies that MR CCSNe like our models can make a sizeable contribution to these isotopes in GCE if they constitute the order of $1\%$ of all CCSNe.

We crudely estimate a quantitative upper bound for the fraction of MR-like CCSN events ($f_{\rm MR}$) with yields similar to our MHD models relative to all CCSNe following the approach of \citet{wanajo11}. Supposing that the isotope with the largest production fraction $^{82}$Se is solely contributed by MR CCSNe and $^{16}$O is contributed by all other CCSNe, we have \footnote{Note that although a single MR CCSN can contribute significant $^{16}$O, $f_{\rm MR}$ is a small number so the overall contribution to $^{16}$O is negligible. }
\begin{equation}
    \frac{f_{\rm MR}}{1-f_{\rm MR}} = \frac{X_{\sun}(^{82}{\rm Se})/X_{\sun}(^{16}{\rm O})}{M_{\rm MR}(^{82}{\rm Se})/\langle M(^{16}{\rm O})\rangle} = 2.8\times10^{-4},
\end{equation}
where we take the solar mass fraction $X_{\sun}$ from \citet{lodders09} and the averaged $^{16}$O production $\langle M(^{16}{\rm O})\rangle$ in CCSNe to be 1.5\,$M_{\sun}$ \citep{wanajo09}. This results in $f_{\rm MR}\sim2.8\times10^{-4}$. Taking the next over-produced isotopes, $^{83}$Kr and $^{76}$Ge, $f_{\rm MR}$ is on the order of $10^{-4}$ to $10^{-3}$. Note that the averaged production of oxygen by CCSNe $\langle M(^{16}{\rm O})\rangle$ is uncertain subject to its dependence on the progenitor mass as well as the integration over the initial mass function of CCSN progenitors. With the theoretical prediction of \citet{sukhbold16} that $\langle M(^{16}{\rm O})\rangle=0.57\,M_{\sun}$, the allowed fraction of MR CCSNe would be even smaller by a factor of $\sim2.6$ than that taking $\langle M(^{16}{\rm O})\rangle=1.5\,M_{\sun}$ as $f_{\rm MR}\propto \langle M(^{16}{\rm O})\rangle$. If without further fallback onto the PNS, the large production factors of these isotopes relative to the solar abundance indicate that the MR CCSNe as studied here should be rare in the history of the Milky Way. Most notably, the limit is lower than the hypernova fraction in the local Universe \citep{smith_11} and far lower than the higher hypernova fractions that may obtain in low-metallicity environments according to transient observations \citep{arcavi_10} and have been invoked to explain GCE at low-metallicity in the past \citep{nomoto_06,kobayashi_06,grimmett_20}. The current simulations clearly cannot represent generic outcomes from hypernova explosions. More simulations of MR CCSNe are needed to determine whether this truly implies a rate constraint on neutron-star forming MR explosions, or whether higher event rates can be made compatible with GCE constraints within modelling uncertainties.

\begin{figure}
    \centering
    \includegraphics[width=0.47\textwidth]{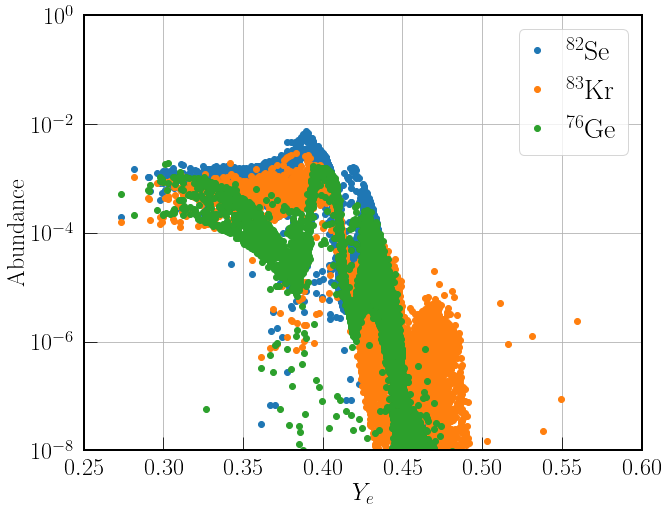}
    \caption{Abundance (mass fraction divided by mass number) of the most overproduced isotopes ($^{82}$Se, $^{83}$Kr and $^{76}$Ge) as a function of the electron fraction $Y_\mathrm{e}$ of tracers in  model m39\_B12.}
    \label{fig:overproduction}
\end{figure}

Because these over-produced isotopes put the most stringent constraint on the rate of MR CCSNe, we further diagnose their origins. In Figure~\ref{fig:overproduction} we plot the abundance of the most over-produced isotopes ($^{82}$Se, $^{83}$Kr, $^{76}$Ge) as a function of $Y_\mathrm{e}$ of tracers in  model m39\_B12. It is clear that these isotopes are tightly correlated and produced by the tracers with $Y_\mathrm{e}\lesssim0.45$. Therefore, their over-production is characteristic of the modestly neutrino-rich ejecta with $Y_\mathrm{e}\in [0.3,0.45]$ in our adopted MR models (cf.~Figure~\ref{fig:mvsye}). 

\section{Conclusions}  \label{sec:con}
We have presented the post-processed nucleosynthesis yields of the innermost ejecta in 3D MR-CCSN simulations of 39\,$M_{\sun}$ progenitor of low metallicity $Z=1/50\,Z_{\sun}$ 
and compared the results with a reference non-magnetic simulation. We analyzed two MR-CCSN models which start from initial magnetic field strengths differing by 2 orders of magnitudes ($10^{10}$ vs. $10^{12}$\,G at the center of the progenitor) but experience very similar explosion dynamics. The MR explosions set in early at $\sim$100\,ms and energetically, reaching an energy of $\sim3\times10^{51}$\,erg due to efficient tapping of the PNS rotational energy reservoir by strong magnetic fields. Magnetically collimated jets are launched after the bulk explosion sets in, and they are not the main driver of the explosion. Our main findings on the nucleosynthesis of MR-CCSNe are as follows:
\begin{itemize}
    \item The bulk ejecta in the MR models have neutron-rich components with $Y_\mathrm{e}$ down to $\sim0.25$, thus allowing the synthesis of weak $r$-process elements until Cs with the mass number up to $\sim$130. However, the adopted MR-CCSN models cannot synthesize heavier elements, i.e. the third $r$-process peak. The ejecta in the non-magnetic model have $Y_\mathrm{e}$ mostly in-between 0.45 and 0.55, and its nucleosynthesis resembles those of other neutrino-driven CCSNe with massive progenitors.
    \item Despite the difference in the initial magnetic field strength, the nucleosynthesis pattern of the two MR-CCSN models are very similar, in accord with their similar explosion dynamics. The most over-produced isotopes ($^{82}$Se, $^{83}$Kr, $^{76}$Ge) put stringent constraints on the fraction of MR-CCSN events like our models relative to all CCSNe. Events like the simulated ones must constitute less than $\sim10^{-3}$ of all CCSNe.
    \item The $^{56}$Ni masses are 0.064\,$M_{\sun}$, 0.194\,$M_{\sun}$, and 0.141\,$M_{\sun}$ in the non-magnetic, weak-magnetic and strong-magnetic models, respectively. The $^{56}$Ni yield and explosion energy put our adopted MR-CCSN models on the low end of the population of observed hypernovae.
\end{itemize} 
In order to draw inferences on supernova explosion mechanisms from abundances of individual stars or stellar populations, it will be essential to first quantify uncertainties of yield predictions by exploring variations with progenitor parameters and model uncertainties. Ultimately this cannot be achieved by a single best set of models, but only by meta-analyses combining yields based on independent simulations by different groups.
To aid this purpose, our nucleosynthesis yields are therefore made publicly available at Zenodo: doi:\href{https://doi.org/10.5281/zenodo.10578981}{10.5281/zenodo.10578981} for further investigation of the impact of MR CCSNe on galactic chemical evolution.

A major uncertainty of the current study is that considerable fallback onto the PNS may occur as the explosion shock propagates out. The binding energy of the material outside the shock is $\sim2\times10^{51}$\,erg as the progenitor is a compact carbon-oxygen star with a radius of $\sim2.7\times10^{10}$\,cm.~\footnote{We note that \citet{powell23} gave a smaller value $\sim10^{51}$\,erg for the binding energy. The difference is that \citet{powell23} only included the material up to the boundary of hydrodynamic simulations ($10^{10}$\,cm), while we have integrated over the whole outer envelope here (up to $\sim2.7\times10^{10}$\,cm).} This overburdened material may even hinder the explosion of this progenitor model without magnetic fields. However, determining the final explosion energy is not just a matter of subtracting the binding energy of the overburden envelope and long-term simulations are indispensable for solving the subtle and highly aspherical outflows and inflows (see e.g. \citealt{chan20,burrows23}). With significant fallback, the rate constraint based on the over-produced isotopes will be alleviated. Long-term simulations for shock propagation and fallback accretion are required to address the genuine chemical contribution of the MR CCSNe.

\begin{acknowledgments}
This work is supported by the National Natural Science Foundation of China (NSFC, Nos. 12288102, 12393811, 12090040/3), the National Key R\&D Program of China (Nos. 2021YFA1600401 and 2021YFA1600400), the International Centre of Supernovae, Yunnan Key Laboratory (No. 202302AN360001), the Natural Science Foundation of Yunnan Province (No. 202201BC070003) and the Yunnan Fundamental Research Project (No. 202401BC070007). The authors gratefully acknowledge the “PHOENIX Supercomputing Platform” jointly operated by the Binary Population Synthesis Group and the Stellar Astrophysics Group at Yunnan Observatories, CAS.
Authors JP and BM are supported by the Australian Research Council (ARC) Centre of Excellence (CoE) for Gravitational Wave Discovery (OzGrav) project number CE170100004.
BM is supported by ARC Future Fellowship FT160100035. JP is supported by the ARC Discovery Early Career Researcher Award (DECRA) project number DE210101050. We acknowledge computer time allocations from Astronomy Australia Limited's ASTAC scheme, the National Computational Merit Allocation Scheme (NCMAS), and from an Australasian Leadership Computing Grant. Some of this work was performed on the Gadi supercomputer with the assistance of resources and services from the National Computational Infrastructure (NCI), which is supported by the Australian Government, and through support by an Australasian Leadership Computing Grant. Some of this work was performed on the OzSTAR national facility at Swinburne University of Technology. OzSTAR is funded by Swinburne University of Technology and the National Collaborative Research Infrastructure Strategy (NCRIS).

\end{acknowledgments}

\vspace{5mm}

\software{ \textsc{CoCoNuT-FMT} \citep{dimmelmeier02,muller15,muller20}; \texttt{SkyNet} \citep{skynet}; \textsc{Numpy} \citep{numpy};  \textsc{Matplotlib} \citep{matplotlib}}

\bibliography{mhd}{}
\bibliographystyle{aasjournal}



\end{document}